\documentstyle[12pt,epsf]{article}
\newcount\eqnum
\def\eqn#1{\global\advance\eqnum by 1
   \xdef#1{ (\the\eqnum ) }(\the\eqnum )  }

\def\be{\begin{equation}}
\def\ee{\end{equation}}
\begin{document}
\title{%
Measuring the Decorrelation Times of Fourier Modes in Simulations}
\author{R. Ben-Av$^1$\\
Physics Department, \\
Princeton University, Princeton, N.J. 08540,USA\\
and\\
G. Bhanot$^2$\\
Thinking Machines Corporation\\
245, First Street, Cambridge, MA 02142, USA\\
and\\
Institute for Advanced Study, Princeton, N.J. 08540, USA}

\maketitle

\begin{abstract}
We describe a method to study the rate at which modes decorrelate in
numerical simulations. We study the XY model updated with the
Metropolis and Wolff dynamics respectively and compute the rate at
which each eigenvector of the dynamics decorrelates. Our method allows
us to identify the decorrelation time for each mode separately. We
find that the autocorrelation function of the various modes is
markedly different for the `local' Metropolis compared to the
`non-local' Wolff dynamics. Equipped with this new insight, it may be
possible to devise highly efficient algorithms.

\vskip 3.5cm
\begin{flushright}
IAS preprint: IASSNS-HEP-93/2\\
E-mail: 1 - radi@puhep1.princeton.edu ; 2 - gyan@think.com
\end{flushright}
\end{abstract}

\newpage
\section{Introduction}
In the last few years there has been a lot of progress in finding new
Monte-Carlo dynamics with little or no critical slowing down. This
progress was initiated by the paper of Swendsen and Wang
\cite{SwendWang} who described a novel cluster dynamics which was not in
the universality class of local dynamics such as the Metropolis or
heat-bath algorithms and had $z<1$ for Potts models.  Their results
were later extended to the $\phi^4$ model \cite{Brower} and to $O(n)$
models \cite{UWolff}. For a recent survey see
\cite{wheretosee}.

However there has not been much progress in understanding what makes
these improved algorithms work. The features that would contribute to
improvement are easy to state: the dynamics must work on `relevant
degrees of freedom'. Unfortunately, these are difficult to isolate for
most models. The Potts models are in a special, simple class where we
expect clusters of spins to constitute the important degrees of
freedom.  However, even here, one does not know why the Swendsen-Wang
algorithm has the dynamical critical exponent that it does. In our
opinion, such an understanding would lead to the invention of better
algorithms.  One should note the work of Lee and Sokal \cite{LiSokal}
which established a lower bound on the dynamical critical exponent $z$
for Ising models.  However, many fundamental questions remain
unanswered, such as: Is there a universality class of `non-local'
dynamics?  What is the mechanism behind the success of acceleration
algorithms?  Does the new dynamics apply to general gauge models,
models with frustrations, and if so, how?

In this paper we take a first step in this direction by describing a
method to study the rate at which modes at different length scales
decorrelate for any given dynamics. We use the $XY$ model in
2-dimensions as a platform to describe our method.  We introduce a
large set of operators and measure their time-time autocorrelation
function for Metropolis and Wolff dynamics. Translation invariance allows us to
identify Fourier modes as eigenvectors of the autocorrelation matrix.
We compute the eigenvalues of the Fourier modes as a function of
time. These directly measure the rate at which the various modes decorrelate.

Our results show that there is a marked difference between the two
methods. In the Metropolis case, high frequency mode converge much
faster than low frequency modes and it is the low frequency modes that
govern the slow decorrelation in this dynamics. For the Wolff update,
all modes decorrelate much faster than the Metropolis update. However,
some high frequency modes of the Wolff dynamics are slower than some
low frequency modes. This latter phenomena, to our knowledge, has
never been demonstrated before.

Because our method provides a direct way to study the rate at which
modes equilibrate, it could be used to both find the slow degrees of
freedom and also to check the effectiveness of different algorithms in
accelerating them. We expect that our method will  help in fine
tuning the search for {\it effective} algorithms.

\section{The Model}

The action of the two dimensional ferromagnetic XY model is:
\be
E = \sum_{<i,j>} cos(\theta_i-\theta_j)
\ee
where $ i=(n_x,n_y), \ \ 1\leq n_x \leq {\rm L_x}$ and
$ 1\leq n_y \leq {\rm L_y}$. The lattice is square with periodic boundary
conditions, $\beta$ is the inverse temperature. This model has the famous
K-T phase transition  \cite{K-T} and has been extensively investigated
numerically  \cite{numXY,Gupta,MGXY}.

The `local' dynamics we use is a parallel Metropolis algorithm. The
dynamics is defined as follows: The lattice is decomposed into a
red-black checkerboard of even and odd sites.  All the even (odd)
spins are updated simultaneously. A sweep is an update of the even
spins followed by an update of the odd spins.  The Wolff dynamics is
implemented in its single cluster version \cite{UWolff}. The origin of
the cluster is chosen randomly and this fact makes the dynamics
translation invariant.  For the moment, we will define one sweep to be
one cluster update.

\section{Autocorrelation Function}

Let us define the elements of the autocorrelation matrix $M(t)$ to be:
\be
M_{i,j} (t) = <cos(\theta_i(0)) cos(\theta_j(t))> - <cos(\theta_i)>^2
\ee
where $i=(n_x,n_y)$ as above.
In various limits, $M_{i,j}(t)$ is related to static quantities.
$M_{i,j}(0)$ is the static correlation function of the observables
$cos(\theta_i)$ and hence it is independent of the dynamics.
$\sum_{i,j} M_{i,j}(0)$ is the magnetic susceptibility times the
volume. $M_{i,j}(\infty)$ is zero.  For $t$ between $0$ and $\infty$,
$M_{i,j}(t)$ depends strongly on the dynamics. It is always easier to analyze
the case where the dynamics is translational invariant. When this is true,
$M_{i,j}(t) = M_{i+d,j+d}(t) $ where $\vec d$ is any lattice vector.
In this case it is obvious that the eigenvectors of
$M_{i,j}$ are the Fourier modes for any $t$.

It is clear that `typewrite order' Metropolis dynamics in two
dimensions is not translationaly invariant \cite{HNprivate}. The
parallel checkerboard Metropolis algorithm we use is translation
invariant for even $\vec k$; that is, for $\vec k = (k_1,k_2)$ such
that both $k_1$ and $k_2$ are even. However, it is easy to define a
set of variables for which the dynamics is translation invariant.
These are block spin variables which are the sums of $cos(\theta_i)$
on the four sites of a $2\times 2$ lattice square. We will compute
$M_{i,j}$ for these variables. The Wolff dynamics we use is obviously
translation invariant. We also use the same block variables for this
dynamics to make the comparison between the two straightforward.

Let us denote the eigenvalue of $M(t)$ by $\lambda_{\vec k}(t)$ where
$\vec k$ is a vector in the first Brillouin zone.  For each $\vec k$
one can assign a decorrelation time $\tau_{\vec k}$ via:
\be
\lambda_{\vec k}(t) \sim  e^{-t/{\tau_{\vec k}}}\quad,t \rightarrow \infty
\ee

\section{The Results}
We have measured the autocorrelation eigenvalues $\lambda_{\vec k}(t)$
for both the Metropolis and Wolff dynamics for lattices with L=16 and
$\beta = 0.81, 0.83, 0.85, 0.87, 0.89, 0.91$. Note that for
$\xi(\beta=0.91) = 20 > L $ \cite{Gupta,MGXY} and so this temperature
is already in the `spin-wave' phase for this lattice size. We did five
sets of $120,000$ sweeps each, discarding the first 20,000. We took
$10,000$ measurements of the autocorrelation function of Eq.2 with ten
sweeps between each measurement. The reason we were limited to small
lattices is that it is quite time consuming to measure the correlation
function for all pairs $i$, $j$ in Eq.~2 (which is what we did).  If
one knows the eigenvectors it is of course possible to measure only
the autocorrelation matrix for them directly. This results in a much
smaller matrix (in our case, it reduces the matrix $M$ from $64\times
64$ to $8\times 8$).  This is the preferred approach for future
analysis.

In Fig.~1a-f we show $\lambda_k$ for both dynamics as a function of
update time $T$ for each $\beta$ value.  The time $T$ is the number of
updates of both colors of the checkerboard in the case of the
Metropolis algorithm.  For the Wolff algorithm, the appropriate
definition for $T$ is the number of cluster updates times the average
cluster size divided by the number of lattice sites. We have only
plotted the data for $T\le 40$. In reality, our measurements extended
out to $T=100$. We plot only the first three modes corresponding to
$\vec k = (0,0)$, $\vec k = (0,1)$ and $\vec k = (0,2)$ where we
measure $\vec k$ in units of $2\pi/L$.  The error bars on the
correlation functions were estimated by repeating each measurement up
to five times and computing the standard deviation.

We fit the data for $\lambda_k$ to Eq.3 in the region where $T$ was
large enough for this relationship to be valid. From these fits, we
extract the autocorrelation time $\tau$ which is shown as a function
of $\beta$ in Figs 2, 3a, and 3b.

\section{Discussion}
First it is clear from Fig.1 that the Wolff algorithm is much superior
to the Metropolis algorithm. All Fourier modes decorrelate faster at
all temperatures for it. It is also evident from Figs.1 and 2 that
for the Wolff algorithm, the zero mode is much faster than the other
modes. This unusual fact can be seen most clearly in Fig.2 where the
$\tau$ for the zero mode is the smallest for all $\beta$ values.  One
also sees in Fig.~1 that the modes exhibit different short time
behavior. In Fig.2 we clearly see another interesting property of
the Wolff dynamics. The autocorrelation time is generally larger for
shorter distance modes. This should be contrasted with the opposite
behavior of the `local-Metropolis' where the shorter distance modes
(modes with higher k) generally decorrelate faster than longer
distance modes (modes with lower k). This pattern is what one would
expect given the fact that the Metropolis algorithm is local and so
should accelerate short distance modes and the Wolff algorithm is more
global and so should do a better job for longer distance modes.
However, the rich pattern that is evident in Figs.2,3 is not
completely intuitive. For example, in Fig3b, one would expect the
$k=(0,3)$ mode to be faster than the $k=(0,1)$ mode but the data
clearly shows that the reverse is true. On the other hand, the relative
order of the modes $(0,2)$ and $(0,1)$ is what one would expect for a
local algorithm.

One might be able to use the information in Figs.~1-3 to arrive at an
algorithm that can take advantage of the best features of the
Metropolis and Wolff dynamics. This would involve doing some number
$n_1$ of Metropolis sweeps followed by some number $n_2$ of Wolff
sweeps. By monitoring the dynamics of a few modes as we have done in
this paper and changing $n_1$ and $n_2$ one would arrive at the
optimum algorithm. One can obviously also add in other methods such as
the multigrid method \cite{MGXY} and the over-relaxation method
\cite{Adler} to get a truly hybrid algorithm.  In fact, it may be that
one need not actually have to adjust the $n_i$'s to find the optimum
combination if the function $\lambda_{\vec k}$ is known for all $t$ and
for all modes. One might then be able to do an analytic calculation to
find the optimum $n_i$'s.

It is well known that there are other degrees of freedom which are
important in the XY model; viz., the non-local vortex excitations
\cite{K-T}. We believe that for a complete treatement of this model
one should include the vortices as well as spins in the definition of
the auto-correlation matrix $M(t)$. In fact, we have tried to do this.
However, we found that the contribution of the vortex-vortex and
spin-vortex part of $M(t)$ was insignificant when compared to the
spin-spin part in Eq.~2. This comes from the fact that the vortex
density is almost everywhere zero and any non-trivial vortex
contribution to $M(t)$ is swamped by the spin contribution. One might
study the vortices by themselves and see what can be learned from them
before attempting to get information about the correlation between the
dynamics of vortices and spins. However, the issue of the relative
normalization of spin and vortex densities remains.

We hope to take up the issues discussed in the last two paragraphs in
a future publication.

{\noindent
{\bf Acknowledgement}

The research of GB was partly supported by U.S.~DOE Grant
DE-FG02-90ER40542 and the Ambrose Monell Foundation. We thank Thinking
Machines Corporation for lots of computer time on their CM-2 and
CM-200 supercomputers.  We thank Herbert Neuberger for discussions and
Steve Adler for a critical reading of the manuscript.}

%
%
%
%
%

\newpage
\centerline{\bf Figure Captions}
\begin{itemize}
\item
{\bf Figure 1a-f:} $\lambda_{\vec k}$ against the time $T$ to
update $L_x\times\L_y$ spins for different temperatures. For each $\vec k$,
the slower dynamics is Metropolis and the faster is Wolff.
\item
{\bf Fig2:} Decorrelation time $\tau$ as a function of $\beta$ for
various $\vec k$ for the Wolff algorithm. Only the error bars on the
least accurate data are shown. Note that the zero mode decorrelates
the fastest (has the smallest $\tau$).
\item
{\bf Fig3a:} $\tau$ versus $\beta$ for the zero mode for the Metropolis
algorithm. Note the relative scale of this figure and Fig.2.

{\bf Fig3b:} $\tau$ versus $\beta$ for some other modes for the Metropolis
algorithm. Again, only one set of error bars are shown to give a
feeling for the accuracy of the data.
\end{itemize}

\newpage
\epsfbox{fig1_a.ps}
\newpage
\epsfbox{fig1_b.ps}
\newpage
\epsfbox{fig1_c.ps}
\newpage
\epsfbox{fig1_d.ps}
\newpage
\epsfbox{fig1_e.ps}
\newpage
\epsfbox{fig1_f.ps}
\newpage
\epsfbox{fig2.ps}
\newpage
\epsfbox{fig3_a.ps}
\newpage
\epsfbox{fig3_b.ps}

\end{document}